\documentclass[12pt]{iopart}

\begin{document}
\title{Calculation of energy levels and transition amplitudes for barium and radium}

\author{V A Dzuba$^1$, and V V Flambaum$^{1,2}$}
\address{$^1$School of Physics, University of New South Wales, Sydney
2052,
Australia\\
$^2$ Physics Division, Argonne National Laboratory, Argonne,
Illinois 60439-4843, USA\\}
\ead{V.A.Dzuba@unsw.edu.au}
\date{\today}

\begin{abstract}

The radium atom is a 
promising system for studying parity and time invariance violating 
weak interactions. 
However, available experimental spectroscopic data for 
radium is insufficient for designing an optimal experimental setup.
We calculate the energy levels and transition amplitudes for radium
states of significant interest. Forty states corresponding to all possible
configurations consisting of the $7s$, $7p$ and $6d$ single-electron states
as well as the states of the $7s8s$, $7s8p$ and $7s7d$ configurations have been
calculated. The energies of ten of these states corresponding to the
$6d^2$, $7s8s$, $7p^2$, and $6d7p$ configurations are not known from
experiment. Calculations for barium are used to control the accuracy.

\end{abstract}

\maketitle

\section{Introduction}

Studying parity ($P$) and time ($T$) invariance violating effects in atoms is a way
of searching for new physics beyond the standard model (see, e.g.~\cite{Ginges}).
These effects are strongly enhanced in radium atom due to high value of the nuclear
charge $Z$, and specific features of the nuclear and electron 
structure~\cite{Auerbach,FlambaumRa,DzubaRa}. The  atomic electric
dipole moment induced by the T,P-violating nuclear forces
 and P-violating effects produced by the nuclear anapole moment
are enhanced 3 orders of magnitude in comparison with previous experiments
(the  detailed comparison and complete list of references may be found  e.g. in
the review  ~\cite{Ginges}).  
Preparations for the measurements are currently in progress at Argonne~\cite{argonne}
and Groningen~\cite{Jungmann,groningen}. 

Detailed knowledge of the positions of the lowest states of an atom as well as
transition probabilities between them is important for the design of cooling and
trapping schemes and for estimation of the enhancement of the $P$ and $T$-odd
effects. Energy spectrum of Ra was first measured by Rasmussen~\cite{Rasmussen}
in 1934. Interpretation of his data was corrected by Russell~\cite{Russell} also
in 1934. Compilation by Moore~\cite{Moore} based on these two works contains
about forty energy levels of radium. There were few more experimental works
on radiums studying Rydberg states~\cite{Armstrong}, hyperfine structure and
isotope shift~\cite{Ahmad, Wendt}, nuclear magnetic moments~\cite{Arnold}, etc.
In the most recent work by Sielzo {\it et al} the lifetime and position
of the $^3$P$^o_1$ state of Ra has been measured. The result for the energy is 
in excellent agreement with early data by Russell~\cite{Russell} and
Rasmussen~\cite{Rasmussen}.


There were some doubts inspired by theoretical work of Biero\'{n} 
{\it et al}\cite{Bieron} among experimentalists working with radium on
whether the data presented by Rasmussen~\cite{Rasmussen} and Russell~\cite{Russell}
were reliable and accurate. This disagreement between theory and experiment
motivated our previous calculations~\cite{DzubaGinges}. The calculations
strongly favored experimental data. However, the strongest evidence
of the correctness of the experimental data came from recent success 
in trapping of radium atoms at Argonne.
Corresponding paper which would include among other things new data
on experimental resolution to the D-state location is to be submitted
soon~\cite{Guest}.

Excellent agreement between theory and experiment for radium and its lighter
analog barium\cite{DzubaGinges} allows us to address next problem - gaps in
experimental data for radium. In particular, it is important to know the 
positions of the states corresponding to the $6d^2$ configuration. 
The locations of these levels are important when considering the
possibility of laser-cooling and trapping Ba or Ra in the metastable
$6s5d$~$^{3}D_{3}$ or $7s6d$~$^{3}D_{3}$ state, respectively.  This
would be a useful alternative to the relatively slow and leaky
transitions available from the ground $^{1}S_{0}$ state.  In
particular, the $6s5d$~$^{3}D_{3}$ - $5d6p$ $^{3}F_{4}$ transition
in barium and the $7s6d$~$^{3}D_{3}$ - $6d7p$ $^{3}F_{4}$ transition
in radium could provide a fast and closed cycling transition.
However, the data
for barium~\cite{Curry} indicates that the energies of the $5d^2$ configuration
lie very low, between the $6s6p$ and the $5d6p$ configuration,
and therefore provide an undesirable leak channel.
It is reasonable
to expect that the energies of the $6d^2$ configuration of radium also lie
pretty low. This would limit
the cooling and trapping schemes for radium causing
leaking of some transitions into the states of the $6d^2$ configuration~\cite{Guest}.

The main task of our previous paper~\cite{DzubaGinges} was to prove that the
experimental data was correct. Therefore we calculated only energy levels  
known from experiment. In present paper we extend the calculations to include
all states of the lowest configurations of radium. We calculate energy levels 
and lifetimes of forty states of the $7s^2$, $7s7p$, $7s6d$, $7s8s$, $6d7p$,
 $6d^2$, $7s8p$, $7p^2$ and $7s7d$ configurations. For 19 lowest states we
also present detailed data on electric dipole transition amplitudes.
Similar calculations for barium are used to control the accuracy of the
calculations.    

\section{Method of calculations and results for barium}

The method of calculations has been described in detail in our previous 
works~\cite{Kozlov96,vn,vn4,DzubaGinges}. Here we repeat its main points 
most relevant to present calculations.

The calculations are done in the $V^{N-2}$ approximation~\cite{vn} which means 
that initial Hartree-Fock procedure is done for a double ionized ion, with
two valence electrons removed. This approach has many advantages.
It simplifies the inclusion of the core-valence correlations by avoiding 
the so called {\it subtraction} diagrams\cite{Kozlov96,vn}. This in turn
allows one to go beyond second-order of the many-body perturbation theory 
(MBPT) in treating core-valence correlations. Inclusion of the higher-order
core-valence correlations significantly improves the accuracy of the 
results~\cite{vn,vn4}. Also, using $V^{N-2}$ approximation makes 
calculations for a positive ion and for a neutral atom very similar
providing more opportunities for the control of the accuracy.
One more advantage is that atomic core is independent on the state
of valence electrons. Ground and excited states are treated equally
which is important for calculating energy intervals.

Single-electron Hamiltonian for a valence electron has the form
\begin{equation}
  \hat h_1 = h_0 + \hat \Sigma_1,
\label{h1}
\end{equation}
where $h_0$ is the relativistic Hartree-Fock Hamiltonian: 
\begin{equation}
  \hat h_0 = c \mathbf{\alpha p} + (\beta -1)mc^2 - \frac{Ze^2}{r} + V^{N-2},
\label{h0}
\end{equation}
and $\hat \Sigma_1$ is the correlation potential operator which represents 
correlation interaction of a valence electron with the core. 
Calculations for a positive ion are done by solving the equation
\begin{equation}
  (\hat h_1 - \epsilon_v)\psi_v = 0,
\label{Brueck}
\end{equation}
where $\epsilon_v$ and $\psi_v$ are the energy and wave function of a 
valence electron. Both $\epsilon_v$ and $\psi_v$ include the effect
of core-valence correlations and the wave functions $\psi_v$ are
often called {\it Brueckner orbitals} to distinguish them from
Hartree-Fock orbitals which do not include correlations.

The effective Hamiltonian for a neutral two-electron atom is the sum
of two single-electron Hamiltonians plus an operator representing
interaction between valence electrons:
\begin{equation}
  \hat H^{\rm eff} = \hat h_1(r_1) + \hat h_1(r_2) + \hat h_2(r_1,r_2).
\label{heff}
\end{equation}
Interaction between valence electrons is the sum of Coulomb interaction
and correlation correction operator $\hat \Sigma_2$:
\begin{equation}
  \hat h_2 = \frac{e^2}{|\mathbf{r_1 - r_2}|} + \hat \Sigma_2(r_1,r_2),
\label{h2}
\end{equation}
$\hat \Sigma_2$ represents screening of Coulomb interaction between 
valence electrons by core electrons.

We use standard configuration interaction (CI) technique to solve the
Schr\"{o}dinger equation for two-electron valence states.
Two-electron wave function for the valence electrons $\Psi$ has a form of 
expansion over single-determinant wave functions
\begin{equation}
  \Psi = \sum_i c_i \Phi_i(r_1,r_2).
\label{psi}
\end{equation}
$\Phi_i$ are constructed from the single-electron valence basis 
states calculated in the $V^{N-2}$ potential
\begin{equation}
 \Phi_i(r_1,r_2) = \frac{1}{\sqrt{2}}(\psi_a(r_1)\psi_b(r_2)-\psi_b(r_1)\psi_a(r_2)).
\label{psiab}
\end{equation}
Coefficients $c_i$ as well as two-electron energies are found by
solving matrix eigenvalue problem
\begin{equation}
  (H^{\rm eff}_{ij} - E)X = 0,
\label{Schr}
\end{equation}
where $H^{\rm eff}_{ij} = \langle \Phi_i | \hat H^{\rm eff} | \Phi_j \rangle$ and
$X = \{c_1,c_2, \dots , c_n \}$.

The most complicated part of the calculations is calculation of the
correlation correction operators $\hat \Sigma_1$ and $\hat \Sigma_2$.
We use MBPT and Feynman diagram technique to do the calculations.
MBPT expansion for $\hat \Sigma$ starts from second order. Inclusion
of the second order operators $\hat \Sigma_1^{(2)}$ and $\hat \Sigma_2^{(2)}$
into effective Hamiltonian (\ref{heff}) accounts for most of the core-valence
correlations. However, further improvement is still possible if higher-order
correlations are included into $\hat \Sigma_1$. We do this the same way as
for a single valence electron atoms~\cite{Dzuba89}. Two dominating
classes of the higher-order diagrams are included into $\hat \Sigma_1$
by applying Feynman diagram technique to the part of $\hat \Sigma_1$
which corresponds to direct Coulomb interaction. These two classes are
(a) screening of Coulomb interaction between valence and core electrons
by other core electrons, and (b) interaction between an electron excited 
from the core and the hole in the core created by this excitation~\cite{Dzuba89}.
The effect of screening of Coulomb interaction in exchange diagrams is 
imitated by introducing screening factors $f_k$ into each Coulomb line.
We assume that screening factors $f_k$ depend only on the multipolarity of
the Coulomb interaction $k$. It turns out that the values of $f_k$ vary
very little from atom to atom and the same values can be used for all
atoms of the first and second columns of the periodic table:
\[ f_0 = 0.72, f_1 = 0.62, f_2 = 0.83, f_3 = 0.89, f_4 = 0.94, f_5 = 1.0, \dots .\]
Calculations show that for atoms like Ba and Ra accurate treatment of
$\hat \Sigma_1$ is more important than that of $\hat \Sigma_2$.
Therefore we calculate $\hat \Sigma_2$ in second order of MBPT only.

One needs a complete set of single-electron states to calculate $\hat \Sigma$
and for construction of two-electron basis states (\ref{psiab}) for the CI
calculations. We use the same basis in both cases. It is constructed using
$B$-spline technique~\cite{Johnson1,Johnson3}. We use 50 $B$-splines of order 7 in a 
cavity of radius $R_{max} = 40 a_B$, where $a_b$ is Bohr radius.
Single-electron basis orbitals in each partial wave are constructed as linear 
combination of 50 $B$-splines
\begin{equation} 
  \psi_a(r) = \sum_{i=1}^{50} b^a_i B_i(r).
\label{bspline}
\end{equation}
Coefficients $b^a_i$ are found from the condition that $\psi_a$ is an
eigenstate of the Hartree-Fork Hamiltonian $h_0$ (\ref{h0}).

The effect of inclusion of second and higher-order $\hat \Sigma$ into
effective Hamiltonian for two-electron valence states of Ba and Ra
was studied in detail in our previous paper~\cite{DzubaGinges}. 
It was also suggested there that the best results can be obtained 
if fitting parameters are introduced before $\hat \Sigma_1$ for each 
partial wave. The values of these parameters for Ba found from fitting
experimental energies of the $6s^2$, $6s6p$ and $6s5d$ configurations are
$\lambda_s = 1.0032$,$\lambda_p = 1.0046$ and $\lambda_d = 0.9164$.
Note that we keep the same fitting parameters for $\hat \Sigma_{p_{1/2}}$ 
and $\hat \Sigma_{p_{3/2}}$ as well as for $\hat \Sigma_{d_{3/2}}$
and $\hat \Sigma_{d_{5/2}}$. We do this to avoid false contribution to 
the fine structure. Fitting of the energies imitate the effects of
higher-order correlations, incompleteness of the basis set, Breit
and QED corrections. 

Final results for Ba are presented in Table \ref{BaEn}. The results for twelve 
states of the $6s^2$, $6s5d$, $6s6p$ and $5d6p$ configurations are the same
as in our previous work. However we present now 27 more states, including 
states of very important $5d^2$ configuration. Note that corresponding
energies absent in Moore book~\cite{Moore} and we use recent compilation
by Curry~\cite{Curry} instead. Parameter $\Delta$ in the Table is the 
difference between experimental and theoretical energies 
($\Delta = E_{expt} - E_{calc}$). The agreement between theory and
experiment is extremely good in most of cases. The largest difference is for
the $5d^2 \ ^1$S$_0$ state. It is 723~cm$^{-1}$ or 2.7\%. Note however
that experimental value for this state came from a different source than
all other data and has the largest uncertainty (see~\cite{Curry} for details).
There is a chance that the experimental value is incorrect. The only other
large difference is for the $5d^2 \ ^1$D$_2$ state. It is 409~cm$^{-1}$ or
1.8\%. For other states of the $5d^2$ configuration the difference 
between theory and experiment is about 1\% or smaller. For most of other
states the difference is just small fraction of a per cent.  

\begin{table}
\caption{\label{BaEn}Energies and lifetimes of lower states of barium}
\begin{tabular}{ccc rrr cccc}
\br Config. & Term & $J$ & \centre{3}{Energies (cm$^{-1}$)} 
 & \centre{3}{$g$-factors} & Lifetime \\ 
 & &  & \centre{1}{Expt\cite{Curry}} & \centre{1}{Calc} & \centre{1}{$\Delta$} 
 & Obs\cite{Curry} & NR & Calc & \\
\mr
$6s^2$    & $^1$S     &  0  &       0.000 &     0 &    0 &      & 0.00 & 0.00 & - \\
$6s5d$    & $^3$D     &  1  &    9033.966 &  9039 &   -5 & 0.53 & 0.50 & 0.50 & - \\
          & $^3$D     &  2  &    9215.501 &  9216 &    0 & 1.18 & 1.17 & 1.16 & - \\
          & $^3$D     &  3  &    9596.533 &  9581 &  -14 & 1.38 & 1.33 & 1.33 & - \\
          & $^1$D     &  2  &   11395.350 & 11626 & -231 & 1.00 & 1.00 & 1.00 & - \\
$6s6p$    & $^3$P$^o$ &  0  &   12266.024 & 12269 &   -3 &      & 0.00 & 0.00 & 2.6 $\mu$s \\
          & $^3$P$^o$ &  1  &   12636.623 & 12637 &    0 & 1.45 & 1.50 & 1.50 & 1.2 $\mu$s \\
          & $^3$P$^o$ &  2  &   13514.745 & 13517 &   -2 & 1.52 & 1.50 & 1.50 & 1.4 $\mu$s \\
          & $^1$P$^o$ &  1  &   18060.261 & 17833 &  227 & 1.02 & 1.00 & 1.00 & 8.2 ns \\
$5d^2$    & $^3$F     &  2  &  20934.035  & 21145 & -211 &      & 0.67 & 0.67 & 190 $\mu$s  \\
          & $^3$F     &  3  &  21250.195  & 21457 & -207 &      & 1.08 & 1.08 & 2.9 ms \\
          & $^3$F     &  4  &  21623.773  & 21831 & -207 &      & 1.25 & 1.25 & - \\

          & $^1$D     &  2  &  23062.051  & 23471 & -409 &      & 1.00 & 1.15 & 470 ns \\
          & $^3$P     &  0  &  23209.048  & 23369 & -160 &      & 0.00 & 0.00 & 160 ns \\
          & $^3$P     &  1  &  23479.976  & 23640 & -160 &      & 1.50 & 1.50 & 170 ns \\
          & $^3$P     &  2  &  23918.915  & 24160 & -241 &      & 1.50 & 1.34 & 270 ns \\
          & $^1$S     &  0  &  26757.3~~  & 26034 &  723 &      & 0.00 & 0.00 & 1.3 $\mu$s \\
$5d6p$    & $^3$F$^o$ &  2  &  22064.645  & 22040 &   25 &      & 0.67 & 0.76 & 33 ns \\
          & $^3$F$^o$ &  3  &  22947.423  & 22926 &   21 &      & 1.08 & 1.08 & 30 ns \\ 
          & $^1$D$^o$ &  2  &  23074.387  & 23078 &   -4 &      & 1.00 & 0.92 & 26 ns \\  
          & $^3$F$^o$ &  4  &  23757.049  & 23745 &   12 &      & 1.25 & 1.25 & 27 ns \\
  
          & $^3$D$^o$ &  1  &  24192.033  & 24149 &   43 & 0.54 & 0.50 & 0.51 & 18 ns \\
          & $^3$D$^o$ &  2  &  24531.513  & 24494 &   38 & 1.16 & 1.17 & 1.17 & 18 ns \\
          & $^3$D$^o$ &  3  &  24979.834  & 24952 &   28 & 1.32 & 1.33 & 1.32 & 18 ns \\

          & $^3$P$^o$ &  0  &  25642.126  & 25705 &  -63 &      & 0.00 & 0.00 & 13 ns \\   
          & $^3$P$^o$ &  1  &  25704.110  & 25765 &  -61 & 1.52 & 1.50 & 1.49 & 13 ns \\
          & $^3$P$^o$ &  2  &  25956.519  & 26022 &  -65 & 1.52 & 1.50 & 1.49 & 14 ns \\
          & $^1$F$^o$ &  3  &  26816.266  & 26881 &  -65 & 1.09 & 1.00 & 1.00 & 47 ns \\
          & $^1$P$^o$ &  1  &  28554.221  & 28604 &  -50 & 1.02 & 1.00 & 1.00 & 14 ns \\
$6s7s$    & $^3$S     &  1  &  26160.293  & 26074 &   86 &      & 2.00 & 2.00 & 16 ns \\
          & $^1$S     &  0  &  28230.231  & 28361 & -131 &      & 0.00 & 0.00 & 29 ns \\
$6s6d$    & $^1$D     &  2  &  30236.826  & 30230 &    7 &      & 1.00 & 1.00 & 38 ns \\
          & $^3$D     &  1  &  30695.617  & 30622 &   73 &      & 0.50 & 0.50 & 14 ns \\
          & $^3$D     &  2  &  30750.672  & 30672 &   79 & 1.11 & 1.17 & 1.16 & 14 ns \\
          & $^3$D     &  3  &  30818.115  & 30731 &   87 & 1.32 & 1.33 & 1.33 & 14 ns \\
$6s7p$    & $^3$P$^o$ &  0  &  30743.490  & 30616 &  127 &      & 0.00 & 0.00 & 110 ns \\
          & $^3$P$^o$ &  1  &  30815.512  & 30686 &  130 &      & 1.50 & 1.50 & 100 ns \\ 
          & $^3$P$^o$ &  2  &  30987.240  & 30856 &  131 &      & 1.50 & 1.50 &  94 ns \\
          & $^1$P$^o$ &  1  &  32547.033  & 32433 &  114 & 1.07 & 1.00 & 1.00 &  12 ns \\
\br
\end{tabular}
\end{table}

Energy levels of barium where calculated by many authors 
before~\cite{Rose78,Migdalek90,Eliav96,Kozlov97,Johnson98,Kozlov99}. The 
scope of the present work does not allow us to cite all these results.
Comprehensive review of previous calculations for Ba is a big task while
our present consideration serves very specific and limited purpose. We just
want to demonstrate that our method work very well for Ba, therefore we can expect
the results of similar quality for Ra which has similar electron structure.

In Table~\ref{BaEn} we also present the values of observed and calculated $g$-factors.
Non-relativistic (NR) values are given by 
\begin{equation}
  g_{NR} = 1 + \frac{J(J+1)-L(L+1)+S(S+1)}{2J(J+1)},
\label{gf}
\end{equation}
where $J$ is total momentum of the atom, $L$ is angular momentum and $S$ is spin.
Comparing calculated values of $g$-factors with observed and non-relativistic values 
is useful for identification of the states.

\begin{table}
\caption{\label{BaE1}Experimental and theoretical transition probabilities for barium}
\begin{tabular}{cccccc}
\br 
                          &                          & Lower & Upper & 
\centre{2}{Transition probability (s$^{-1}$)} \\
$\lambda_{\rm air}/\AA$ 
& $\Delta E/{\rm cm}^{-1}$ & level & level & Expt.\cite{Curry} &
Calc. \\
\mr
5535.481 & 18060.261 & $6s^2 \ ^1$S$_0$ & $6s6p \ ^1$P$_1$ & $1.19 \times 10^8$ & $1.21 \times 10^8$ \\
5826.274 & 17158.872 & $6s5d \ ^1$D$_2$ & $5d6p \ ^1$P$_1$ & $4.50 \times 10^7$ & $4.14 \times 10^7$ \\
6527.311 & 15316.012 & $6s5d \ ^3$D$_2$ & $5d6p \ ^3$D$_2$ & $3.30 \times 10^7$ & $3.08 \times 10^7$ \\
6595.325 & 15158.068 & $6s5d \ ^3$D$_1$ & $5d6p \ ^3$D$_1$ & $3.80 \times 10^7$ & $3.64 \times 10^7$ \\
6675.270 & 14976.532 & $6s5d \ ^3$D$_2$ & $5d6p \ ^3$D$_1$ & $1.89 \times 10^7$ & $1.67 \times 10^7$ \\
6693.842 & 14934.980 & $6s5d \ ^3$D$_3$ & $5d6p \ ^3$D$_2$ & $1.46 \times 10^7$ & $1.26 \times 10^7$ \\
\br
\end{tabular}
\end{table}

In Table~\ref{BaEn} we also present calculated values of lifetimes of all
considered states. Only electric dipole (E1) transitions were included in the
calculations. Therefore, we don't present lifetimes of the long living states
which can only decay via magnetic dipole (M1) or electric quadrupole (E2)
transitions. 

We calculate E1 transition amplitudes between states $\Psi_a$ and $\Psi_b$ 
using the expression
\begin{equation}
  A(E1)_{ab} = \sum_{i,j} c^a_i c^b_j \langle \Phi_i || d_z + \delta \hat V^{N-2}|| 
\Phi_j \rangle,
\label{E1}
\end{equation}
where $\mathbf{d} = -e\mathbf{r}$ is the electric dipole operator,
$\delta \hat V^{N-2}$ is the correction to the self-consistent potential
of the atomic core due to the electric field of the photon. The term
with $\delta \hat V^{N-2}$ accounts for the so called RPA (random-phase approximation)
or core polarization correction. The functions $\Psi_i$ are two-electron basis
states~(\ref{psiab}) and $c_i$ are expansion coefficients for states $\Psi_a$ and
$\Psi_b$ over basis states $\Phi_i$ as in (\ref{psi}).

Expression (\ref{E1}) is approximate. It includes dominating contributions to the 
$E1$ amplitudes but doesn't take into account some small corrections. A detailed
discussion of different contributions into matrix elements between many-electron
wave functions can be found e.g. in Ref.~\cite{Kozlov98}. In terms of that paper
expression (\ref{E1}) corresponds to the leading contribution to the effective
amplitude ($A_{\rm RPA}$, see Eq.~(22) of Ref.~\cite{Kozlov98}). It accounts for
configuration interaction, core-valence correlations and core polarization effects.
Next, the so called {\it subtraction} contribution ($A_{\rm SBT}$) does not
exist in present calculations since we use the $V^{N-2}$ approximation. Subtraction
terms appear only if Hartree-Fock procedure includes valence electrons. They
account for the difference between Hartree-Fock potential and potential of
the core in the CI Hamiltonian. In the $V^{N-2}$ approximation for a two valence
electrons atom these two potentials are identical. The terms not included into (\ref{E1}) are:
the two-particle correction ($A_{\rm TP}$),the self-energy correction ($A_{\sigma}$),
structure radiation and normalization corrections (see \cite{Kozlov98} for details).

Using expression (\ref{E1}) gives satisfactory accuracy for most of the cases.
However neglecting other contributions for small amplitudes may lead to some
instability of the results. This is especially true for small amplitudes which vanish
in the non-relativistic limit ($\Delta S>0, \Delta L>1$). Present calculations give
only rough estimation of the values of these amplitudes. However, it doesn't have 
much effect on lifetimes since lifetimes dominate by strong transitions with
large amplitudes.

Typical accuracy of the calculations for strong transitions is
illustrated by the data in Table~\ref{BaE1}. Here we compare some 
calculated transition probabilities for Ba with the most accurate experimental data.
The probability of the E1 transition from state $i$ to a lower state $j$ is
(atomic units)
\begin{equation}
  T_{ij} = \frac{4}{3} (\alpha \omega_{ij})^3 \frac{A_{ij}^2}{2J_i+1}.
\label{T1}
\end{equation}

\section{Results for radium}

\begin{table}
\caption{\label{RaEn}Energies and lifetimes of lower states of radium}
\begin{tabular}{ccc rrrrr ccc}
\br Config. & Term & $J$ & \centre{5}{Energies (cm$^{-1}$)} 
 & \centre{2}{$g$-factors} & Lifetime \\ 
 & &  & \centre{1}{Expt\cite{Curry}} & \centre{1}{Calc} & \centre{1}{$\Delta$} 
 & \centre{1}{Extrap.} & \centre{1}{$\Delta$} 
 & NR & Calc & \\
\mr
$7s^2$    & $^1$S     &  0  &      0.00  &     0 &       &     0 &       & 0.00 & 0.00 & - \\
					 		 			
$7s7p$    & $^3$P$^o$ &  0  &  13078.44  & 13102 &   -24 & 13099 &   -21 & 0.00 & 0.00 & -  \\
          & $^3$P$^o$ &  1  &  13999.38  & 14001 &    -2 & 14002 &    -2 & 1.50 & 1.47 & 360 ns  \\
          & $^3$P$^o$ &  2  &  16688.54  & 16698 &    -9 & 16696 &    -7 & 1.50 & 1.50 & 5.4 $\mu$s  \\

$7s6d$    & $^3$D     &  1  &  13715.85  & 13742 &   -26 & 13737 &   -21 & 0.50 & 0.50 & 640 $\mu$s \\
          & $^3$D     &  2  &  13993.97  & 13994 &     0 & 13994 &     0 & 1.17 & 1.16 & - \\
          & $^3$D     &  3  &  14707.35  & 14655 &    52 & 14641 &    66 & 1.33 & 1.33 & - \\

$7s6d$    & $^1$D     &  2  &  17081.45  & 17343 &  -262 & 17112 &   -31 & 1.00 & 1.01 & 710 $\mu$s \\

$7s7p$    & $^1$P$^o$ &  1  &  20715.71  & 20433 &   283 & 20660 &    56 & 1.00 & 1.02 & 5.5 ns \\

$7s8s$    & $^3$S     &  1  &  26754.05  & 26665 &    89 & 26751 &     3 & 2.00 & 2.00 & 18 ns \\
$7s8s$    & $^1$S     &  0  &            & 27768 &       & 27637 &       & 0.00 & 0.00 & 80 ns \\

$6d7p$    & $^3$F$^o$ &  2  &  28038.05  & 27991 &    47 & 28016 &    22 & 0.67 & 0.74 & 33 ns \\
          & $^3$F$^o$ &  3  &  30117.78  & 30067 &    51 & 30088 &    30 & 1.08 & 1.09 & 28 ns \\ 
          & $^3$F$^o$ &  4  &  32367.78  & 32363 &     5 & 32375 &    -7 & 1.25 & 1.25 & 23 ns \\  

$6d7p$    & $^1$D$^o$ &  2  &  30918.14  & 30894 &    24 & 30890 &    28 & 1.00 & 1.07 & 19 ns \\  

$6d^2$    & $^3$F     &  2  &            & 29731 &       & 29520 &       & 0.67 & 0.71 & 1.6 $\mu$s  \\
          & $^3$F     &  3  &            & 30464 &       & 30257 &       & 1.08 & 1.08 & 34 $\mu$s \\
          & $^3$F     &  4  &            & 31172 &       & 30965 &       & 1.25 & 1.25 & 3 s \\

$6d^2$    & $^1$D     &  2  &            & 30982 &       & 30573 &       & 1.00 & 1.05 & 150 ns \\

$7s8p$    & $^3$P$^o$ &  0  &  31085.88  & 31008 &    78 & 31135 &    49 & 0.00 & 0.00 & 76 ns \\
          & $^3$P$^o$ &  1  &  31563.29  & 30695 &   868 &       &       & 1.50 & 1.07 & 20 ns \\ 
          & $^3$P$^o$ &  2  &  31874.44  & 31778 &    96 & 31909 &    35 & 1.50 & 1.44 & 57 ns \\

$7p^2$    & $^3$P     &  0  &            & 29840 &       &       &       &      &      & 21 ns \\
          & $^3$P     &  1  &  31248.61  & 31365 &  -116 &       &       & 1.50 & 1.49 & 26 ns \\ 
          & $^3$P     &  2  &  32941.13  & 33180 &  -239 &       &       & 1.50 & 1.21 & 42 ns \\

$7s7d$    & $^3$D     &  1  &  32000.82  & 31895 &   106 & 31968 &    33 & 0.50 & 0.51 & 18 ns \\
          & $^3$D     &  2  &  31993.41  & 31902 &    91 & 31981 &    12 & 1.17 & 1.16 & 19 ns \\
          & $^3$D     &  3  &  32197.28  & 32068 &   129 & 32155 &    42 & 1.33 & 1.33 & 21 ns \\

$7p^2$    & $^1$D     &  2  &  32214.84  & 32205 &    10 &       &       & 1.00 & 1.20 & 29 ns \\

$6d7p$    & $^3$D$^o$ &  1  &  32229.97  & 32090 &   140 &       &       & 0.50 & 0.84 & 21 ns \\
          & $^3$D$^o$ &  2  &  32506.59  & 32436 &    71 &       &       & 1.17 & 1.17 & 13 ns \\
          & $^3$D$^o$ &  3  &  33197.46  & 33169 &    28 &       &       & 1.33 & 1.17 & 21 ns \\

$7s8p$    & $^1$P$^o$ &  1  &  32857.68  & 31446 &  1412 &       &       & 1.00 & 1.16 & 34 ns \\

$6d7p$    & $^3$P$^o$ &  0  &  33782.41  & 33809 &   -27 &       &       & 0.00 & 0.00 & 10 ns \\   
          & $^3$P$^o$ &  1  &  33823.70  & 33837 &   -13 &       &       & 1.50 & 1.40 & 10 ns \\
          & $^3$P$^o$ &  2  &  34382.91  & 34421 &   -38 &       &       & 1.50 & 1.42 & 11 ns \\

$6d^2$    & $^1$S     &  0  &            & 33961 &       &       &       & 0.00 & 0.00 & 150 ns \\ 

$6d7p$    & $^1$F$^o$ &  3  &            & 34332 &       &       &       & 1.00 & 1.14 & 25 ns \\

$6d^2$    & $^1$S     &  0  &            & 35408 &       &       &       & 0.00 & 0.00 & 30 ns \\

$6d7p$    & $^1$P$^o$ &  1  &            & 36043 &       &       &       & 1.00 & 1.03 & 38 ns \\

\br
\end{tabular}
\end{table}

The results of calculations for energies, $g$-factors and lifetimes of forty lowest states 
of radium are presented in Table~\ref{RaEn}. Energies are compared with available
experimental data.
Calculations follow the same procedure as for barium. The only difference is
in values of rescaling parameters for correlation potential $\hat \Sigma$.
Fitting of the experimental energies of the $7s^2$, $7s7p$ and $7s6d$ configurations 
leads to the following values of the rescaling parameters:
$\lambda_s = 1.0021$,$\lambda_p = 1.0053$ and $\lambda_d = 0.9327$.
These values are very close to similar values for barium (see above).
The Coulomb integrals and correlation corrections in electronic analogues
(e.g. Ba and Ra) usually have approximately the same values.
 This fact may be used to extract unaccounted higher correlation corrections
from Ba and  improve
our predictions for unknown energy levels in Ra.  
Indeed, the differences between theory 
and experiment for similar states of radium and barium are very close
at least for lower states. This is
in spite of different order of levels, about 2.5 times difference in fine structure
intervals (spin-orbit interaction increases $\sim Z^2$) and some
difference in fitting parameters for the correlation potential.
This means that the difference between theory and experiment for barium
can be used to improve the predicted positions of those states of radium for
which experimental data is absent. Column {\it Extrap.} in Table~\ref{RaEn}
presents energies of radium corrected using the difference between theory
and experiment for barium. States where experimental data is available
illustrate that the procedure leads to systematic improvement of the
agreement between theory and experiment for lower states of radium.
For states where there is no experimental data extrapolated values give
better prediction of the energies than just {\it ab initio} calculations.

Note that this procedure doesn't work for higher states. This is because
saturation of the basis in the CI calculations rapidly
deteriorates with the increase of the excitation energy. Since the energies 
of similar excited configurations of Ba and Ra are significantly different
the effect of incompleteness of the basis is different too. 

Experimental data for $g$-factors of radium is not available. However,
comparing calculated and non-relativistic values of $g$-factors indicates
that the $L-S$ scheme still works very well for the most of the lower states
of Ra and can be unambiguously used to name the states. The $L-S$
scheme breaks higher in the spectrum due to the combination of relativistic
effects and configuration mixing. For example, as can be seen from Table~\ref{RaEn}
states $7s8p \ ^3$P$^o_1$, $6d7p \ ^3$D$^o_1$ and $7s8p \ ^1$P$^o_1$ are strongly mixed. 
The calculated $g$-factors of each of these states deviate significantly
from the non-relativistic values. This makes it difficult to identify the states.
Also, strong configuration mixing is probably the reason for poor 
agreement between theory and experiment for the energies of these states.

\begin{table*}
\caption{\label{E1amp}E1-transition amplitudes for 19 lowest states of radium}
\begin{tabular}{llll c llll c}
\br
\centre{2}{Even state} & \centre{2}{Odd state} & \centre{1}{Amplitude} &
\centre{2}{Even state} & \centre{2}{Odd state} & \centre{1}{Amplitude} \\
\mr
$7s^2$ & $^1$S$_0$ & $7s7p$ & $^3$P$^o_1$ & 1.218 & $7s6d$ & $^1$D$_2$ & $6d7p$ & $^1$D$^o_2$ & 5.704 \\
       &           & $7s7p$ & $^1$P$^o_1$ & 5.504 &        &           & $6d7p$ & $^3$F$^o_3$ & 0.774 \\
$7s8s$ & $^1$S$_0$ & $7s7p$ & $^3$P$^o_1$ & 0.057 & $6d^2$ & $^3$F$_2$ & $7s7p$ & $^3$P$^o_1$ & 0.542 \\
       &           & $7s7p$ & $^1$P$^o_1$ & 4.176 &        &           & $7s7p$ & $^1$P$^o_1$ & 0.442 \\
$7s6d$ & $^3$D$_1$ & $7s7p$ & $^3$P$^o_0$ & 2.995 &        &           & $7s7p$ & $^3$P$^o_2$ & 0.266 \\
       &           & $7s7p$ & $^3$P$^o_1$ & 2.574 &        &           & $6d7p$ & $^3$F$^o_2$ & 4.644 \\
       &           & $7s7p$ & $^1$P$^o_1$ & 0.437 &        &           & $6d7p$ & $^1$D$^o_2$ & 1.208 \\
       &           & $7s7p$ & $^3$P$^o_2$ & 0.688 &        &           & $6d7p$ & $^3$F$^o_3$ & 1.786 \\
       &           & $6d7p$ & $^3$F$^o_2$ & 3.729 & $6d^2$ & $^1$D$_2$ & $7s7p$ & $^3$P$^o_1$ & 1.274 \\
       &           & $6d7p$ & $^1$D$^o_2$ & 1.394 &        &           & $7s7p$ & $^1$P$^o_1$ & 1.023 \\
$7s8s$ & $^3$S$_1$ & $7s7p$ & $^3$P$^o_0$ & 2.214 &        &           & $7s7p$ & $^3$P$^o_2$ & 1.535 \\
       &           & $7s7p$ & $^3$P$^o_1$ & 3.890 &        &           & $6d7p$ & $^3$F$^o_2$ & 0.290 \\
       &           & $7s7p$ & $^1$P$^o_1$ & 1.476 &        &           & $6d7p$ & $^1$D$^o_2$ & 3.259 \\
       &           & $7s7p$ & $^3$P$^o_2$ & 6.075 &        &           & $6d7p$ & $^3$F$^o_3$ & 0.595 \\
       &           & $6d7p$ & $^3$F$^o_2$ & 0.266 & $7s6d$ & $^3$D$_3$ & $7s7p$ & $^3$P$^o_2$ & 6.340 \\
       &           & $6d7p$ & $^1$D$^o_2$ & 3.584 &        &           & $6d7p$ & $^3$F$^o_2$ & 0.107 \\
$7s6d$ & $^3$D$_2$ & $7s7p$ & $^3$P$^o_1$ & 4.382 &        &           & $6d7p$ & $^1$D$^o_2$ & 2.911 \\
       &           & $7s7p$ & $^1$P$^o_1$ & 0.813 &        &           & $6d7p$ & $^3$F$^o_3$ & 3.064 \\
       &           & $7s7p$ & $^3$P$^o_2$ & 2.605 &        &           & $6d7p$ & $^3$F$^o_4$ & 5.885 \\
       &           & $6d7p$ & $^3$F$^o_2$ & 2.946 & $6d^2$ & $^3$F$_3$ & $7s7p$ & $^3$P$^o_2$ & 0.190 \\
       &           & $6d7p$ & $^1$D$^o_2$ & 0.168 &        &           & $6d7p$ & $^3$F$^o_2$ & 0.702 \\
       &           & $6d7p$ & $^3$F$^o_3$ & 4.568 &        &           & $6d7p$ & $^1$D$^o_2$ & 0.566 \\
$7s6d$ & $^1$D$_2$ & $7s7p$ & $^3$P$^o_1$ & 0.344 &        &           & $6d7p$ & $^3$F$^o_3$ & 5.672 \\
       &           & $7s7p$ & $^1$P$^o_1$ & 3.189 &        &           & $6d7p$ & $^3$F$^o_4$ & 1.597 \\
       &           & $7s7p$ & $^3$P$^o_2$ & 0.510 & $6d^2$ & $^3$F$_4$ & $6d7p$ & $^3$F$^o_3$ & 0.037 \\
       &           & $6d7p$ & $^3$F$^o_2$ & 2.856 &        &           & $6d7p$ & $^3$F$^o_4$ & 6.343 \\
\br
\end{tabular}
\end{table*}

Lifetimes of the states presented in Table~\ref{RaEn} were calculated using 
Eqs.~(\ref{E1}) and (\ref{T1}) for all possible electric dipole transitions
from a given state to lower states. This involves 270 E1-transition amplitudes.
It is impractical to present all of them in a table. However, for considering
different trapping and cooling schemes it is important to know transition
probabilities between different pairs of states rather than just lifetimes.
Therefore we present in Table~\ref{E1amp} 52 amplitudes between 19 lowest
states of radium. This data should be sufficient in most of cases.
More data is available from authors on request. Note that the values of 
small amplitudes which vanish in non-relativistic limit ($\Delta S>0, \Delta L>1$)
should be considered as rough estimation only (see discussion in previous section). 

\ack{}

The authors are grateful to Jeffrey Guest and Zheng-Tian Lu for many stimulating
discussions. 
 This work was supported by the Australian Research Council.
        One of us (VF) appreciates support from Department of Energy,
        Office of Nuclear Physics, Contract No.  W-31-109-ENG-38.

\end{document}